\documentclass[a4paper]{jpconf}
\usepackage{amssymb,bm,graphicx}
\usepackage{amsmath}
\usepackage{amsfonts}
\usepackage{amssymb}
\usepackage{graphicx}%
\providecommand{\U}[1]{\protect\rule{.1in}{.1in}}
\begin{document}
\title{The polarized TMDs in the covariant\\ parton model approach%
\footnote{Contribution to the Proceedings of the 19th International Spin Physics Symposium (SPIN2010), J\"{u}lich, Germany, September 27 - October 2, 2010}}
\author{A.V.~Efremov$^{1}$, P. Schweitzer$^{2}$, O.~V.~Teryaev$^{1}$, \underline{P.~Zavada}$^{3}$}
\address{$^{1}$ {Bogoliubov Laboratory of Theoretical Physics, JINR, 141980 Dubna, Russia}}
\address{$^{2}$ {Department of Physics, University of Connecticut, Storrs, CT 06269, U.S.A.}}
\address{$^{3}$ {Institute of Physics AS CR, Na Slovance 2, CZ-182 21 Prague 8, Czech Rep.}}

\ead{zavada@fzu.cz}

\begin{abstract}
We derive relations between polarized transverse momentum dependent
distribution functions (TMDs) and the usual parton distribution functions
(PDFs) in the 3D covariant parton model, which follow from Lorentz invariance
and the assumption of a rotationally symmetric distribution of parton momenta
in the nucleon rest frame. Using the known PDF $g_{1}^{q}(x)$ as input we
predict the $x$- and $\mathbf{p}_{T}$-dependence of all polarized twist-2
naively time-reversal even (T-even) TMDs.

\end{abstract}

\label{sec1}

TMDs \cite{tmds,Mulders:1995dh} open a new way to a more complete
understanding of the quark-gluon structure of the nucleon. Indeed, some
experimental observations can hardly be explained without a more accurate and
realistic 3D picture of the nucleon, which naturally includes transverse
motion. The azimuthal asymmetry in the distribution of hadrons produced in
deep-inelastic lepton-nucleon scattering (DIS), known as the Cahn effect
\cite{Cahn:1978se}, is a classical example. The intrinsic (transversal) parton
motion is also crucial for the explanation of some spin effects
\cite{Airapetian:1999tv}--\cite{Adams:2003fx}.

{}

In previous studies we discussed the covariant parton model, which is based on
the 3D picture of parton momenta with rotational symmetry in the nucleon rest
frame
\cite{Zavada:1996kp}--\cite{Efremov:2010cy}.

In this model we studied all
T-even TMDs and derived a set of relations among them \cite{Efremov:2009ze}.
It should be remarked that some of the relations among different TMDs were
found (sometimes before) also in other models
\cite{Jakob:1997wg}--\cite{Pasquini:2010pa}.

In the recent paper \cite{Efremov:2010mt} we further develop and broadly
extend our studies
\cite{Zavada:2009sk}--\cite{Efremov:2009vb}
of the relations
between TMDs and PDFs. The formulation of the model in terms of the light-cone
formalism \cite{Efremov:2009ze} allows us to compute the leading-twist TMDs by
means of the light-front correlators $\phi(x,\mathbf{p}_{T})_{ij}$
\cite{Mulders:1995dh} as:%
\begin{align}
\frac{1}{2}\,\mathrm{tr}\left[  \gamma^{+}\;\phi(x,\mathbf{p}_{T})\right]   &
=f_{1}^{q}(x,\mathbf{p}_{T})-\frac{\varepsilon^{jk}p_{T}^{j}S_{T}^{k}}%
{M}\,f_{1T}^{\perp q}(x,\mathbf{p}_{T}),\label{e1}\\
\frac{1}{2}\,\mathrm{tr}\left[  \gamma^{+}\gamma_{5}\phi(x,\mathbf{p}%
_{T})\right]   &  =S_{L}g_{1}^{q}(x,\mathbf{p}_{T})+\frac{\mathbf{p}%
_{T}\mathbf{S}_{T}}{M}g_{1T}^{\bot q}(x,\mathbf{p}_{T}),\label{e2}\\
\hspace{-8mm}\frac{1}{2}\,\mathrm{tr}\left[  i\sigma^{j+}\gamma_{5}%
\phi(x,\mathbf{p}_{T})\right]   &  =S_{T}^{j}\,h_{1}^{q}(x,\mathbf{p}%
_{T})+S_{L}\,\frac{p_{T}^{j}}{M}\,h_{1L}^{\perp q}(x,\mathbf{p}_{T}%
)\label{ee2}\\
&  +\frac{(p_{T}^{j}p_{T}^{k}-\frac{1}{2}\,\mathbf{p}_{T}^{2}\delta^{jk}%
)S_{T}^{k}}{M^{2}}\,h_{1T}^{\perp q}(x,\mathbf{p}_{T})+\frac{\varepsilon
^{jk}p_{T}^{k}}{M}\,h_{1}^{\perp q}(x,\mathbf{p}_{T}).\;\;\nonumber
\end{align}
In the present contribution we report about new results related to the
polarized distributions \cite{Efremov:2010mt}.

\begin{figure}[ptb]
\begin{center}
\includegraphics[width=12cm]{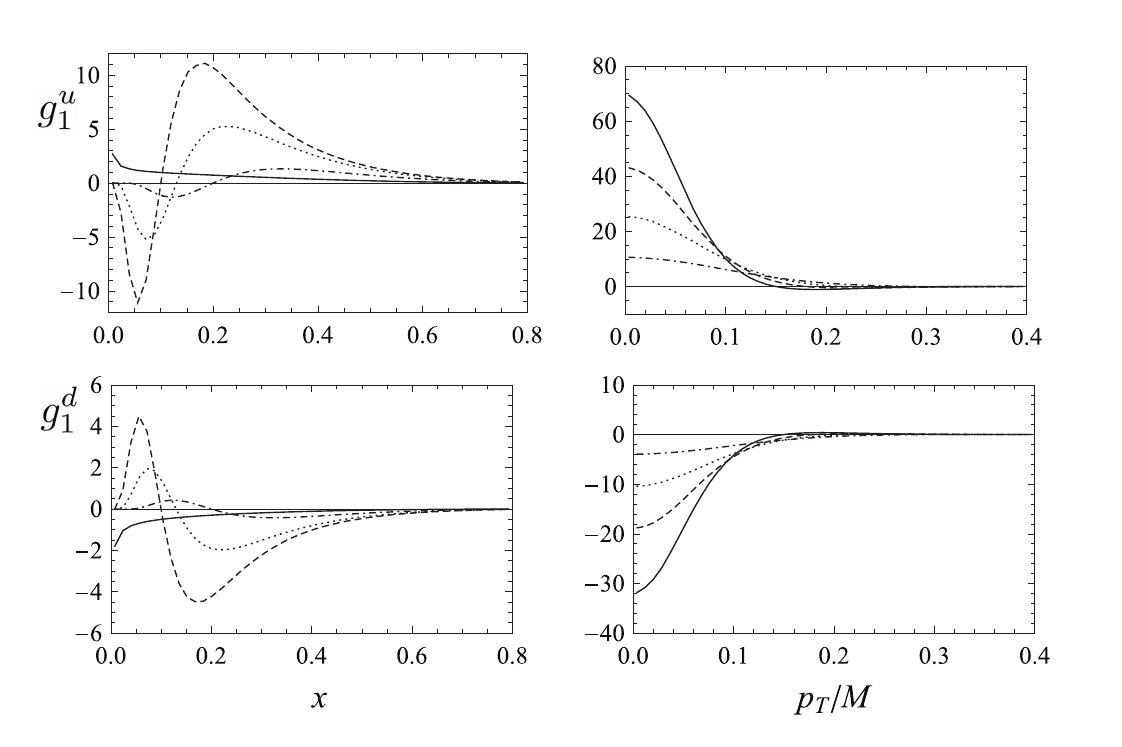}
\par
\vspace{-3mm}\caption{The TMD $g_{1}^{q}(x,\mathbf{p_{T}})$ for $u$-
(\textit{upper panel}) and $d$-quarks (\textit{lower panel}). \textbf{Left
panel}: $g_{1}^{q}(x,\mathbf{p_{T}})$ as function of $x$ for $p_{T}/M=0.10$
(dashed), 0.13 (dotted), 0.20 (dash-dotted line). The solid line corresponds
to the input distribution $g_{1}^{q}(x)$. \textbf{Right panel}: $g_{1}%
^{q}(x,\mathbf{p_{T}})$ as function of $p_{T}/M$ for $x=0.15$ (solid), 0.18
(dashed), 0.22 (dotted), 0.30 (dash-dotted line).}%
\label{ff3}%
\end{center}
\end{figure}

In our approach all polarized leading-twist T-even TMDs are described in terms
of the \textsl{same} polarized covariant 3D distribution $H(p^{0})$. This
follows from the compliance of the approach with relations following from QCD
equations of motion \cite{Efremov:2009ze}. As a consequence all polarized TMDs
can be expressed in terms a single \textquotedblleft generating
function\textquotedblright\ $K^{q}(x,\mathbf{p}_{T})$ as follows
\begin{equation}
\renewcommand{\arraystretch}{2.2}%
\begin{array}
[c]{rcrcl}%
g_{1}^{q}(x,\mathbf{p}_{T}) & = & \displaystyle\frac{1}{2x}\left(  \left(
x+\frac{m}{M}\right)  ^{2}-\frac{\mathbf{p}_{T}^{2}}{M^{2}}\right)  & \times &
K^{q}(x,\mathbf{p}_{T})\;,\\
h_{1}^{q}(x,\mathbf{p}_{T}) & = & \displaystyle\frac{1}{2x}\left(  x+\frac
{m}{M}\right)  ^{2} & \times & K^{q}(x,\mathbf{p}_{T})\;,\\
g_{1T}^{\perp q}(x,\mathbf{p}_{T}) & = & \displaystyle\frac{1}{x}\left(
x+\frac{m}{M}\right)  \; & \times & K^{q}(x,\mathbf{p}_{T})\;,\\
h_{1L}^{\perp q}(x,\mathbf{p}_{T}) & = & \displaystyle-\,\frac{1}{x}\left(
x+\frac{m}{M}\right)  \; & \times & K^{q}(x,\mathbf{p}_{T})\;,\\
h_{1T}^{\perp q}(x,\mathbf{p}_{T}) & = & \displaystyle-\frac{1}{x}\, & \times
& K^{q}(x,\mathbf{p}_{T})\;.
\end{array}
\label{Eq:all-TMDs}%
\end{equation}
with the \textquotedblleft generating function\textquotedblright%
\ $K^{q}(x,\mathbf{p}_{T})$ defined (in the compact notation
of \cite{Efremov:2009ze}) by
\begin{equation}
K^{q}(x,\mathbf{p}_{T})=M^{2}x\int\mathrm{d}\{p^{1}\}\;\;,\;\;\;\;\mathrm{d}%
\{p^{1}\}\equiv\frac{\mathrm{d}p^{1}}{p^{0}}\;\frac{H^{q}(p^{0})}{p^{0}%
+m}\;\delta\left(  \frac{p^{0}-p^{1}}{M}-x\right)  \,.
\label{Eq:generating-function}%
\end{equation}
We have shown that due to rotational symmetry the following relations hold:
\begin{equation}
K^{q}(x,\mathbf{p}_{T})=M^{2}\frac{H^{q}(\bar{p}^{0})}{\bar{p}^{0}+m}%
,\qquad\bar{p}^{0}=\frac{1}{2}\,xM\,\left(  1+\frac{\mathbf{p}_{T}^{2}+m^{2}%
}{x^{2}M^{2}}\right)  , \label{Eq:generating-function-2}%
\end{equation}

\begin{equation}
\pi x^{2}M^{3}H^{q}\!\left(  \frac{M}{2}x\right)  =2\int_{x}^{1}%
\frac{\mathrm{d}y}{y}\;g_{1}^{q}(y)+3\,g_{1}^{q}(x)-x\;\frac{\mathrm{d}%
g_{1}^{q}(x)}{\mathrm{d}x}, \label{Eq:relation-g1-Hp}%
\end{equation}
where we took the limit $m\rightarrow0$ in (\ref{Eq:relation-g1-Hp}). In that
limit we obtain for the generating function (\ref{Eq:generating-function-2})
the result
\begin{equation}
K^{q}(x,\mathbf{p}_{T})=\frac{H^{q}(\frac{M}{2}\xi)}{\,\frac{M}{2}\xi}%
=\frac{2}{\pi\xi^{3}M^{4}}\left(  2\int_{\xi}^{1}\frac{\mathrm{d}y}{y}%
\;g_{1}^{q}(y)+3\,g_{1}^{q}(\xi)-x\;\frac{\mathrm{d}g_{1}^{q}(\xi)}%
{\mathrm{d}\xi}\right)  ,\quad\xi=\,x\,\left(  1+\frac{\mathbf{p}_{T}^{2}%
}{x^{2}M^{2}}\right)  . \label{e3}%
\end{equation}
and from (\ref{Eq:all-TMDs}) we obtain
\begin{equation}
g_{1}^{q}(x,\mathbf{p}_{T})=\frac{2x-\xi}{\pi\xi^{3}M^{3}}\left(  2\int_{\xi
}^{1}\frac{\mathrm{d}y}{y}\;g_{1}^{q}(y)+3\,g_{1}^{q}(\xi)-\xi\;\frac
{\mathrm{d}g_{1}^{q}(\xi)}{\mathrm{d}\xi}\right)  . \label{e4}%
\end{equation}
This relation yields for $g_{1}^{q}(x,\mathbf{p}_{T})$, with the LO
parameterization of \cite{lss} for $g_{1}^{q}(x)$ at $4\,\mathrm{GeV}^{2}$,
the results shown in Fig.~\ref{ff3}.

\begin{figure}[ptb]
\begin{center}
\includegraphics[width=10.7cm]{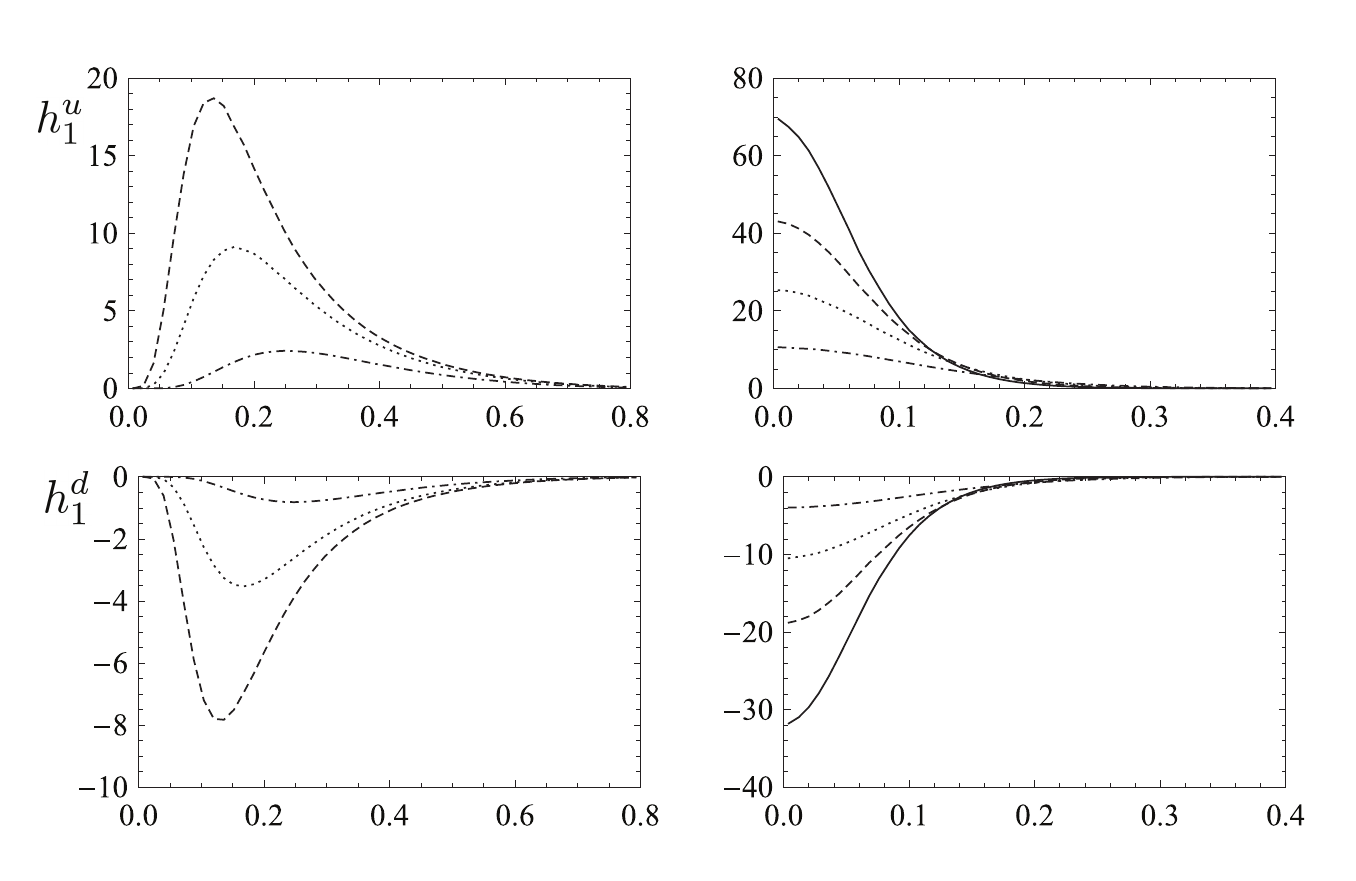}\\
\includegraphics[width=10.7cm]{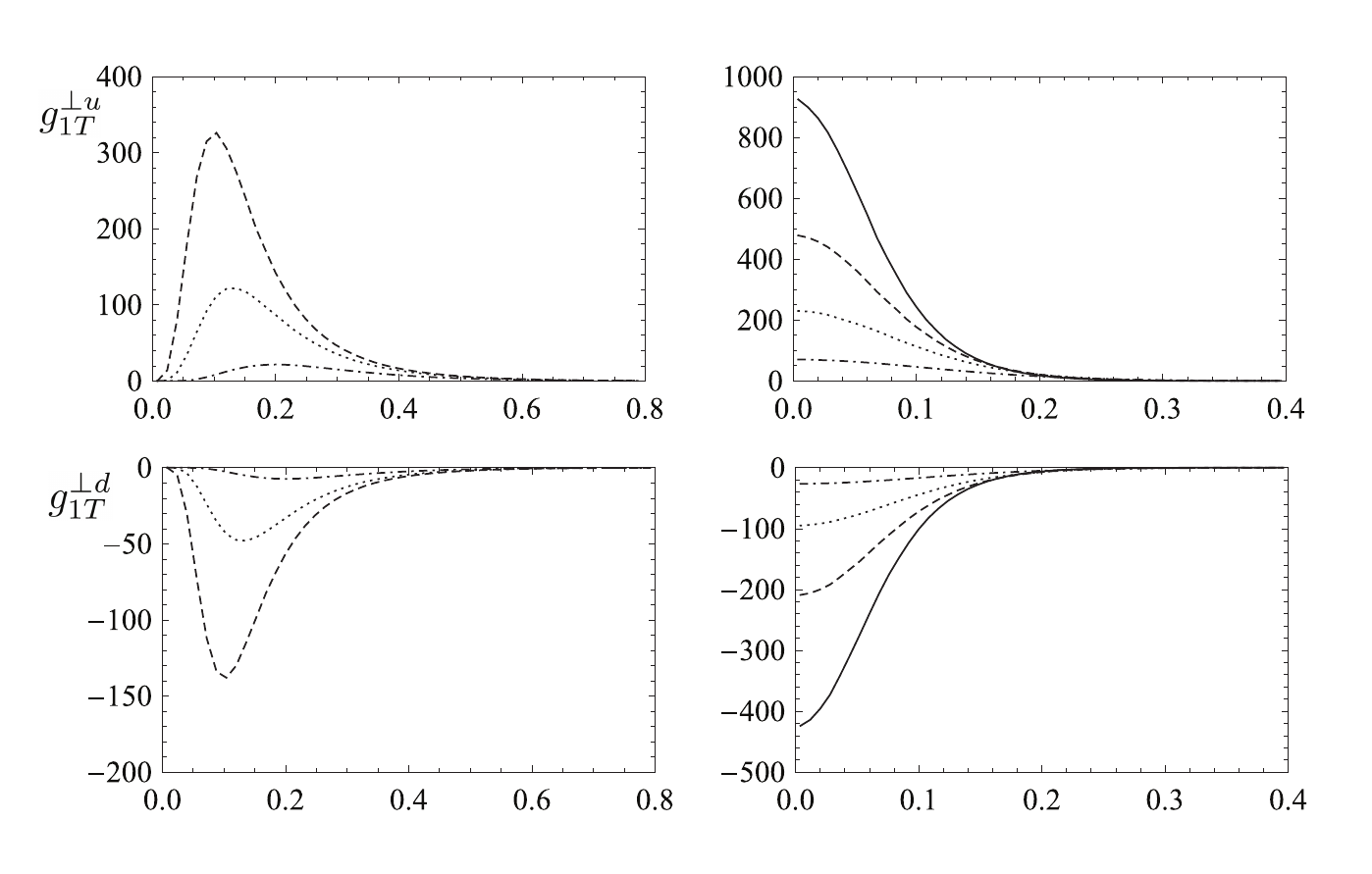}\\
\includegraphics[width=10.7cm]{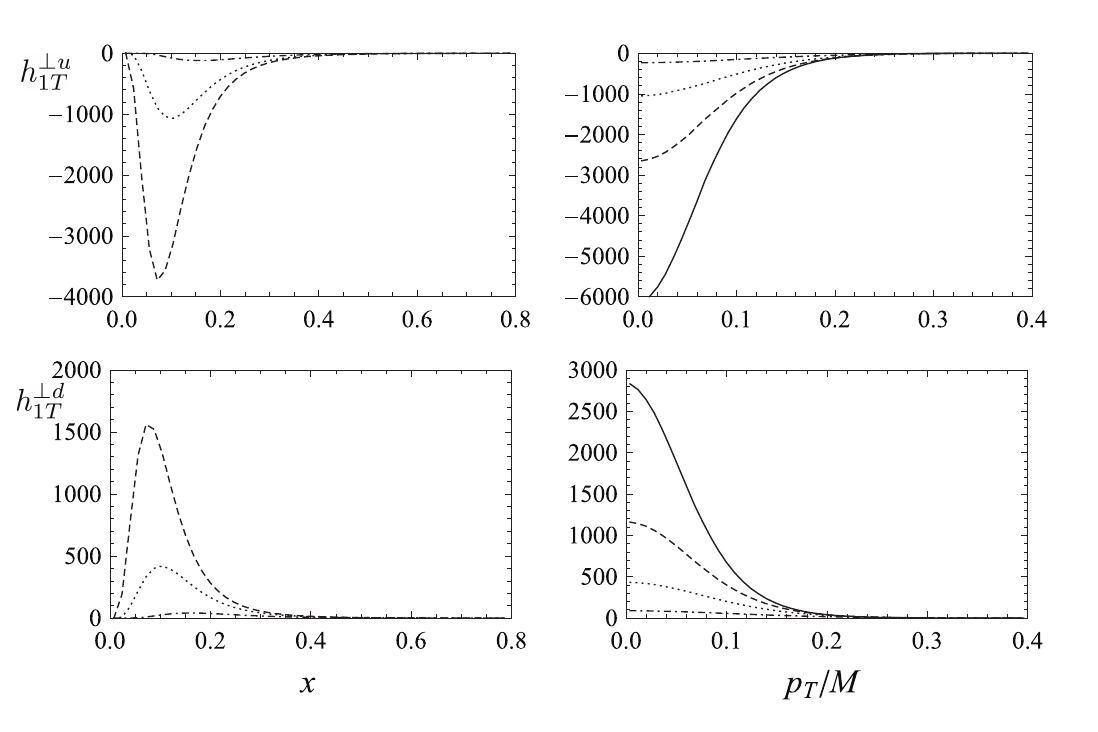} \vspace{-3mm} \caption{The TMDs
$h_{1}^{q}(x,\mathbf{p_{T}})$, $g_{1T}^{\bot q}(x,\mathbf{p_{T}})$,
$h_{1T}^{\perp q}(x,\mathbf{p_{T}})$ for $u$- and $d$-quarks. \textbf{\ Left
panel}: The TMDs as functions of $x$ for $p_{T}/M=0.10$ (dashed), 0.13
(dotted), 0.20(dash-dotted lines). \textbf{Right panel}: The TMDs as functions
of $p_{T}/M$ for $x=0.15$ (solid), 0.18 (dashed), 0.22 (dotted), 0.30
(dash-dotted lines).}%
\label{ff4}%
\end{center}
\end{figure}

The remarkable observation is that $g_{1}^{q}(x,\mathbf{p}_{T})$
changes sign at the point $p_{T}=Mx$, which is due to the prefactor (this is
the definition of the variable $\bar{p}^{1}$ in the limit $m\rightarrow0$)
\begin{equation}
2x-\xi=x\left(  1-\left(  \frac{p_{T}}{Mx}\right)  ^{2}\right)  =-2\bar{p}%
^{1}/M \label{m23}%
\end{equation}
in (\ref{e4}). The expression in (\ref{m23}) is proportional to the quark
longitudinal momentum $\bar{p}^{1}$ in the proton rest frame, which is
determined by $x$ and $p_{T}$ \cite{Efremov:2010mt}. This means, that the sign
of $g_{1}^{q}(x,p_{T})$ is controlled by sign of $\bar{p}^{1}$. To observe
these dramatic sign changes one may look for multi-hadron jet-like final
states in SIDIS. Performing the cutoff for transverse momenta from below and
from above, respectively, should affect the sign of asymmetry.

There is some similarity to $g_{2}^{q}(x)$ which also changes sign, and is
given in the model by \cite{Zavada:2007ww}
\begin{equation}
g_{2}^{q}(x)=\frac{1}{2}\int H^{q}(p^{0})\left(  p^{1}-\frac{\left(
p^{1}\right)  ^{2}-p_{T}^{2}/2}{p^{0}+m}\right)  \:\delta\left(  \frac
{p^{0}-p^{1}}{M}-x\right)  \frac{d^{3}p}{p^{0}}. \label{sp11}%
\end{equation}
The $\delta-$function implies that, for our choice of the light-cone
direction, large $x$ are correlated with large and negative $p^{1}$, while low
$x$ are correlated with large and positive $p^{1}$. Thus, $g_{2}(x)$\ changes
sign, because the integrand in (\ref{sp11}) changes sign between the extreme
values of $p^{1}$. Let us remark, that the calculation of $g_{2}(x)$ based on
the relation (\ref{sp11}) well agrees \cite{Zavada:2002uz} with the
experimental data.

The other TMDs (\ref{Eq:all-TMDs}) can be calculated similarly and differ, in
the limit $m\to0$, by simple $x$-dependent prefactors
\begin{equation}
h_{1}^{q}(x,\mathbf{p}_{T}) =\frac{x}{2}K^{q}(x,\mathbf{p}_{T}), \;\;
g_{1T}^{\perp q}(x,\mathbf{p}_{T}) = K^{q}(x,\mathbf{p}_{T}),\;\;
h_{1T}^{\perp q}(x,\mathbf{p}_{T}) =-\frac{1}{x}K^{q}(x,\mathbf{p}%
_{T}).\label{e5}%
\end{equation}
The resulting plots are shown in Fig.~\ref{ff4}. We do not plot $h_{1L}^{\perp
q}$ since this TMD is equal to $-g_{1T}^{\perp q}$ in our approach
\cite{Efremov:2009ze}. Let us remark, that $g_{1}^{q}(x,\mathbf{p}_{T})$ is
the only TMD which can change sign. The other TMDs have all definite signs,
which follows from (\ref{Eq:all-TMDs},~\ref{e5}). Note also that pretzelosity
$h_{1T}^{\perp q}(x,\mathbf{p}_{T})$, due to the prefactor $1/x$, has the
largest absolute value among all TMDs. Noteworthy, pretzelosity is related to
quark orbital angular momentum in some quark models
\cite{She:2009jq,Avakian:2010br}, including the present approach
\cite{Efremov:2010cy}.

To conclude, let us remark that an experimental check of the predicted TMDs
requires care. In fact, TMDs are not directly measurable quantities unlike
structure functions. What one can measure for instance in semi-inclusive DIS
is a convolution with a quark fragmentation function. This naturally
\textquotedblleft dilutes\textquotedblright\ the effects of TMDs, and makes it
difficult to observe for instance the prominent sign change in the helicity
distribution, see Fig. \ref{ff3}. A dedicated study of the phenomenological
implications of our results is in progress.

\vspace{1mm}

\noindent\textbf{Acknowledgements.} A.~E. and O.~T. are supported by the
Grants RFBR 09-02-01149 and 09-02-00732, and (also P.Z.) Votruba-Blokhitsev
Programs of JINR. P.~Z. is supported by the project AV0Z10100502 of the
Academy of Sciences of the Czech Republic. The work was supported in part by
DOE contract DE-AC05-06OR23177. We would like to thank also Jacques Soffer and
Claude Bourrely for helpful comments on an earlier stage of this study.

\end{document}